\documentclass[%
preprint,
amsmath,amssymb,
prl,
longbibliography,
superscriptaddress,
]{revtex4-1}

\usepackage{graphicx}
\usepackage{dcolumn}
\usepackage{bm}

\begin{document}

\title{Single Pulse Manipulations in Synthetic Time-frequency Space}

\author{Guangzhen Li}
\affiliation{State Key Laboratory of Advanced Optical Communication Systems and Networks, School of Physics and Astronomy, Shanghai Jiao Tong University, Shanghai 200240, China}

\author{Danying Yu}
\affiliation{State Key Laboratory of Advanced Optical Communication Systems and Networks, School of Physics and Astronomy, Shanghai Jiao Tong University, Shanghai 200240, China}

\author{Luqi Yuan}
\email{yuanluqi@sjtu.edu.cn}
\affiliation{State Key Laboratory of Advanced Optical Communication Systems and Networks, School of Physics and Astronomy, Shanghai Jiao Tong University, Shanghai 200240, China}

\author{Xianfeng Chen}
\affiliation{State Key Laboratory of Advanced Optical Communication Systems and Networks, School of Physics and Astronomy, Shanghai Jiao Tong University, Shanghai 200240, China}
\affiliation{Shanghai Research Center for Quantum Sciences, Shanghai 201315, China}
\affiliation{Jinan Institute of Quantum Technology, Jinan 250101, China}
\affiliation{Collaborative Innovation Center of Light Manipulation and Applications, Shandong Normal University, Jinan 250358, China}


\begin{abstract}
Synthetic dimensions in photonic structures provide unique opportunities for actively manipulating light in multiple degrees of freedom. Here, we theoretically explore a dispersive waveguide under the dynamic phase modulation that supports single pulse manipulations in the synthetic (2+1) dimensions. Compared with the counterpart of the conventional (2+1) space-time, we explore temporal diffraction and frequency conversion in a synthetic time-frequency space while the pulse evolves along the spatial dimension. By introducing the effective gauge potential well for photons in the synthetic time-frequency space with the control of the modulation phase, we show that a rich set of pulse propagation behaviors can be achieved, including  confined pulse propagation, fast/slow light, and pulse compression. With the additional  nonlinear oscillation subject to the effective  force along the frequency axis of light, we provide an exotic approach for actively manipulating the single pulse in both temporal and spectral domains, which shows the great promise for  applications  of the pulse processing and optical communications in integrated photonics.
\end{abstract}

\maketitle


Synthetic dimension in photonics is an emergent field  for  exploring physics in higher-dimensional space within  lower-geometrical structure, which  also points towards manipulating light by utilizing physical phenomena in the synthetic space \cite{Yuan:18,ozawa2019topological}. Different degrees of freedom of photons can be used to construct  synthetic dimensions, such as frequencies \cite{yuan2016photonic,ozawa2016synthetic,Bell:17,qin2018spectrum},  orbital angular momenta \cite{luo2015quantum},  pulse arrival times \cite{PhysRevLett.107.233902,regensburger2012parity}, and others \cite{lustig2019photonic,maczewsky2020synthesizing,wang2020multidimensional}.   With synthetic dimensions, many potential applications have been proposed, including  unidirectional frequency translation \cite{Yuan:161}, orbital angular momentum switch \cite{PhysRevA.97.043841}, pulse narrowing \cite{doi:10.1063/1.5039375}, and mode-locked topological laser \cite{PhysRevX.10.011059}.
Furthermore,  it has also been shown that one can simultaneously build up two synthetic dimensions with different degrees of freedom of light and explore topological edge states \cite{yuan2019photonic,dutt2020a}. This synthetic (2+1) dimensions require only one single cavity, which  dramatically simplify the experimental requirements.

Group velocity dispersion (GVD) is a fundamental optical characteristic in a medium, and is of great importance  in ultrashort pulse manipulations \cite{reeves2003transformation,PhysRevLett.114.053901,divitt2019ultrafast}, such as generation of  optical solitons \cite{PhysRevLett.95.143902,lee2017towards}, pulse compression \cite{PhysRevLett.79.4566,colman2010temporal,tan2010monolithic}, and group velocity control \cite{gehring2006observation,Li:15,qin2020fast}, where the interplay between dispersive and nonlinear effects on optical pulses  take place \cite{agrawal2000nonlinear}. Moreover, it has been found that, when a pulse propagates along a dispersive waveguide, one can consider the problem in a synthetic (1+1) dimensions, i.e., the optical field diffracts along a time dimension when it evolves along the spatial dimension \cite{peschel2008discreteness,PhysRevLett.115.183901,Plansinis:16}.

\begin{figure}[htb]
\center
\includegraphics[width=8.0 cm]{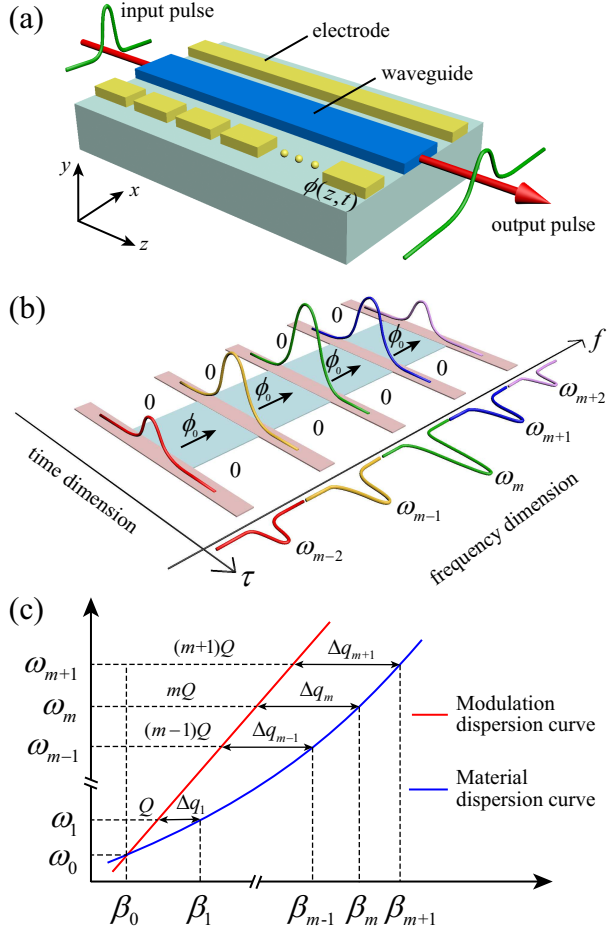}
\caption{\label{fig1} (a) A  pulse propagating through a waveguide with segmented electrodes for  modulations. (b) The system in (a) can be mapped into a  synthetic (2+1) dimensions, where an effective gauge potential well is constructed in the time-frequency space  while the pulse propagates along $z$. (c) Dispersion curves for waveguide (blue line) and  modulation (red line). }
\end{figure}

In this work, we move a step further and show the possibility of multiple single pulse manipulations in a synthetic (2+1) dimensions including the temporal diffraction and frequency conversion in a synthetic time-frequency space while a pulse propagates along the spatial dimension. A dispersive  waveguide incorporating segmented electrodes under travelling wave electro-optic  modulation is considered [see Fig.~\ref{fig1}(a)]. We show that one construct a two-dimensional synthetic space including the time and frequency dimensions, and pulse dynamics is studied when the spatial propagating dimension is treated as the synthetic time evolution. An effective gauge potential well is constructed in synthetic two dimensions with non-uniform distribution of modulation phases to confine light \cite{PhysRevX.4.031031,lumer2019light,Cohen2019light}. By manipulating the effective  well in multiple ways, we show rich physics of pulse manipulations, including confined pulse propagation, fast/slow light, and pulse compression. Fundamentally different from previous works \cite{peschel2008discreteness,PhysRevLett.115.183901,Plansinis:16}, our results link to physics in a (2+1) dimensions, which points out exotic route towards manipulating pulse profile and frequency conversion process. Our work  can find important applications of optical pulse engineering in various platforms, ranging from second/third-order dispersive waveguide-based systems to  on-chip dispersive microresonator-based systems.

We begin with briefly illustrating  synthetic (2+1) dimensions  constructed in a waveguide  shown in Fig. 1(a), modulated by  a traveling wave with a sinusoidal radio frequency (RF) signal. The  refractive index is governed by \cite{gan1997traveling}
\begin{equation}
  n(z,t)=n_0+\Delta n\cos[\Omega t-Qz+\phi (z,t)],
\end{equation}
where $n_0$ is static refractive index and $\Delta n$ is the modulation amplitude. $\Omega$, $Q$, and $\phi$  are  the frequency, wavevector, and modulation phase of the RF signal. For a pulse centered at the frequency $\omega_0$ propagating along the spatial dimension $z$, the applied  modulation connects field components at discrete frequencies  $\omega_m=\omega_0+m\Omega$, and forms the  synthetic  frequency dimension [see  Fig.~\ref{fig1}(b)] \cite{qin2018spectrum}.
On the other hand, for a dispersive waveguide, the pulse experiences temporal diffraction, which brings up the concept of the synthetic dimension along the continuous retarded time frame ($\tau = t- z/v_g$, with $v_g$ being the group velocity at $\omega_0$) \cite{peschel2008discreteness,agrawal2000nonlinear}. Hence, a synthetic continuous-discrete time-frequency space is constructed for the pulse travelling along $z$-direction inside the waveguide.

We then consider a time-dependent and spatially non-uniform modulation phase $\phi(z,\tau)$, which can be achieved by controlling RF signals at each segmented electrode. Such phase distribution supports the effective gauge potential for light in the synthetic space \cite{fang2012realizing}. In particular, we take the form of  $\phi(z,\tau)=\phi_0$ as
\begin{equation}
\phi(z,\tau)=\left\{
\begin{aligned}
& \phi_0 & ~~|\tau-\tau_\texttt{c}(z)|\leq \Delta\tau_{\phi}(z), \\
& 0   & ~~|\tau-\tau_\texttt{c}(z)|>\Delta\tau_{\phi}(z),
\end{aligned}
\right.
\end{equation}
i.e., at any $z$, the hopping phase along the frequency dimension is $\phi_0$  in a region  with the center  $\tau_\texttt{c}(z)$ and   width $\Delta\tau_{\phi}(z)$,  and equals to 0 at the remaining regions. Such phase distribution brings an \textit{effective gauge potential well} in the synthetic space with the size dependent on $z$ [see Fig.~1(b)]  \cite{PhysRevX.4.031031,lumer2019light,Cohen2019light}, and can be used to manipulate light in different ways.

We now build detailed theoretical framework to study the system. For  pulse propagating through the modulated waveguide, the electric field of the pulse can be expanded as $E(z,t)=\sum_{m}a_m(z,t)e^{i(\omega_mt-\beta_mz)}$, where $a_m(z,t)$ is the slowly-varying envelope for the frequency component at $\omega_m$ \cite{agrawal2000nonlinear,boyd2020nonlinear}. The propagation constant $\beta_m$ is not equally spaced due to GVD  [see Fig.~\ref{fig1}(c)], which can be defined as $\beta_m=\beta_0+mQ+\Delta q_m$. Here, $\beta_0$ is the wavevector associated with $\omega_0$, and $\Delta q_m$ donates the  wavevector mismatching.  The pulse field follows the wave equation \cite{boyd2020nonlinear}
\begin{equation}\label{eq1}
  \frac{\partial^2E(z,t)}{\partial z^2}-\frac{1}{\varepsilon_0c^2}\frac{\partial^2[\varepsilon_0\varepsilon_r(z,t)E(z,t)]}{\partial t^2}=0,
\end{equation}
where $\varepsilon_0$  and $\varepsilon_r(z,t)=n^2(z,t)$ are vacuum and relative permittivity, respectively. With the expansion of the field, we obtain the propagating equations for $m^{\texttt{th}}$ component in the retarded frame  \cite{supplementary}
\begin{equation}\label{eq6}
  i\frac{\partial a_m(z,\tau)}{\partial z}= -\frac{k_2}{2}\frac{\partial^2a_m}{\partial \tau^2}+  g[a_{m+1}e^{-i(c_1+c_2+2c_2m)z-i\phi(z,\tau)}+a_{m-1}e^{i(c_1-c_2+2c_2m)z+i\phi(z,\tau)}].
 \end{equation}
Here,  $c_1=k_1\Omega-Q$ is  linear  mismatching between light and RF signal, and $c_2=k_2\Omega^2/2$  is  nonlinear mismatching caused by GVD, where $k_1$ and $k_2$ are  Taylor expansion coefficiencies of wavevector $k(\omega)$ around $\omega_m$,  representing the reciprocal of the group velocity and GVD, respectively. $g=\Delta n\omega_0/2c$ donates the coupling strength.

Equation~(\ref{eq6}) describes the dynamics of a pulse with multiple frequency components at $\omega_m$ propagating along $z$-axis  in a reference frame moving at $v_g$. The first term on right-band side of Eq.~(\ref{eq6})  dominates the pulse dispersion behavior, which is a counterpart of  wave diffusion in the spatial dimension, while the second term  refers  frequency conversions. Equation~(\ref{eq6}) is therefore  the nonlinear Schr\"{o}dinger equation within $(2+1)$ dimensions \cite{agrawal2000nonlinear,PhysRevA.95.033801}, where the pulse experiences continuous temporal diffraction and discrete frequency conversion in two synthetic dimensions while it evolves at $z$-axis [see Fig.~\ref{fig1}(b)]. Moreover, the hopping phase in Eq.~(\ref{eq6}) gives an effective nonlinear force $F=c_1+2c_2m+\partial\phi/\partial z$ pointing along the  frequency dimension \cite{Yuan:161}. Different from previous studies, which have explored consequences of an effective linear force in the synthetic  frequency dimension \cite{Yuan:161,qin2018spectrum,Lieabe4335},  here the effective force is nonlinear due to  GVD. Yet, as we  show in the following, a synthetic time-frequency space together with the nonlinear force give us alternative opportunity for manipulating the pulse propagation in the waveguide.

\begin{figure}[htb]
\center
\includegraphics[width=8.0 cm]{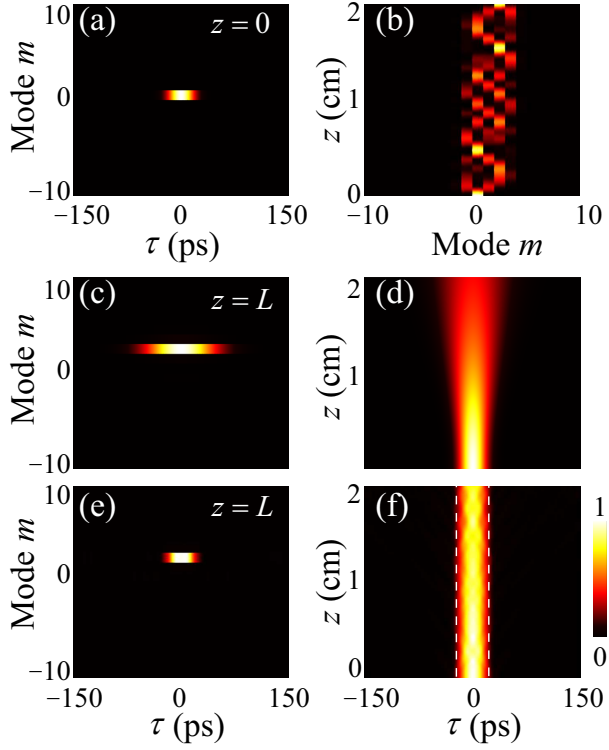}
\caption{\label{fig2}  Pulse propagations under (b)-(d) constant  modulation and (e)-(f) modulation with effective gauge potential well labeled by dashed line.  (a), (c) and (e) Intensity distribution $|a_m(z,\tau)|^2$ at  $z=0$ and $z=L$ in the synthetic time-frequency space.  (b)  Evolution of intensities for each mode $I_m(z)\equiv\int_{\tau}|a_m(z,\tau)|^2\texttt{d}\tau$. (d) and (f) Evolution of the pulse $I(z,\tau)\equiv\sum_{m}|a_m(z,\tau)|^2$. }
\end{figure}

We simulate pulse propagations  by Eq.~(\ref{eq6}) with excitations at one end of the waveguide at $z=0$. The input pulse has a profile as $f(\tau) =e^{-1.386[\tau/\Delta\tau]^2}$, where $\Delta\tau=30$ ps is the  temporal full width at half maximum (FWHM). We assume that the input field contains only one frequency component  at $\omega_0=1.2\times10^{15}$ Hz  (or 1550 nm). Note that Eq.~(\ref{eq6}) is valid when the condition $\Omega>\Delta\omega(z)$ is satisfied, i.e., field profiles at different frequency components in the spectral domain  do not overlap, where $\Delta\omega(z)$ is the spectral FWHM for each frequency component at any $z$. For the input pulse, $\Delta\omega=2\pi\cdot14.7$ GHz. The simulation  performed with 21 modes ($m=-10,...,10$). Figure~\ref{fig2}(a) shows the profile of the input pulse in the synthetic time-frequency space.

First, we consider the normal case that pulse propagates in the waveguide under the modulation with $\phi(z,\tau)=0$. We choose modulation with $\Delta n=5\times10^{-4}$ and $\Omega=2\pi\cdot29.4$ GHz, which gives $g=10^3$ m$^{-1}$. For the  waveguide with a length $L=2$ cm, we have $k_2=4\times10^{-20}$ m$^{-1}$s$^2$, which lead to $c_2=682$ m$^{-1}$.  $c_1$  can be approximated to be zero by tuning $Q=k_1\Omega$. All parameters are chosen with the experimental feasibility in the literature \cite{luennemann2003electrooptic,8721132,wang2018integrated}.

When the pulse evolves along the $z$ axis, it experiences frequency conversion and nonlinear  oscillation near the $0^{\texttt{th}}$ mode in the  frequency dimension under the effective nonlinear force [see Fig.~\ref{fig2}(b)]. At  $L=2$ cm, frequency components of pulse oscillate back to the single mode, which shifts to the $2^{\texttt{nd}}$ mode [see Fig. 2(c)]. Moreover, the pulse gets broadened due to GVD and has the temporal width $\Delta\tau(z=L)=79$ ps, as shown in Figs.~\ref{fig2}(c)-(d). It agrees well with the theoretical calculated evolution of the pulse width for a  Gaussian pulse [see Fig.~S1(a)] \cite{supplementary}.

We next consider modulations with non-uniform distributions of phases and explore the dynamics from Eq.~(\ref{eq6}) under effective gauge potential well. We choose parameters $\phi_0=\pi$, $\tau_\texttt{c}(z)=0$ and $\Delta\tau_{\phi}(z)=30$ ps  which  indicates a fixed well in the synthetic space. The simulation shows the confinement of light in middle region  of the effective well (labelled by the dashed line) in the synthetic time dimension as shown in Figs.~\ref{fig2}(e)-(f),  while the frequency component is shifted to the $2^{\texttt{nd}}$ mode as previous case. The pulse width remains 30 ps while maintaining an approximate  Gaussian waveform during the propagation [see Fig. S1(b)] \cite{supplementary}. The result here shows an interesting combination between nonlinear  oscillation along the frequency dimension and the light confinement due to the effective gauge potential well in the time dimension. As a comparison, we calculate the evolution of pulse with $c_2=0$, with other parameters unchanged, and find temporal confinement of light  persists while the frequency conversion covers a broad range of modes \cite{supplementary}. Not only the confined pulse propagation demonstrated here, the idea of utilizing effective gauge potentials in synthetic time-frequency dimensions provides more different ways to manipulate the pulse.

\begin{figure}[htb]
\center
\includegraphics[width=8.0 cm]{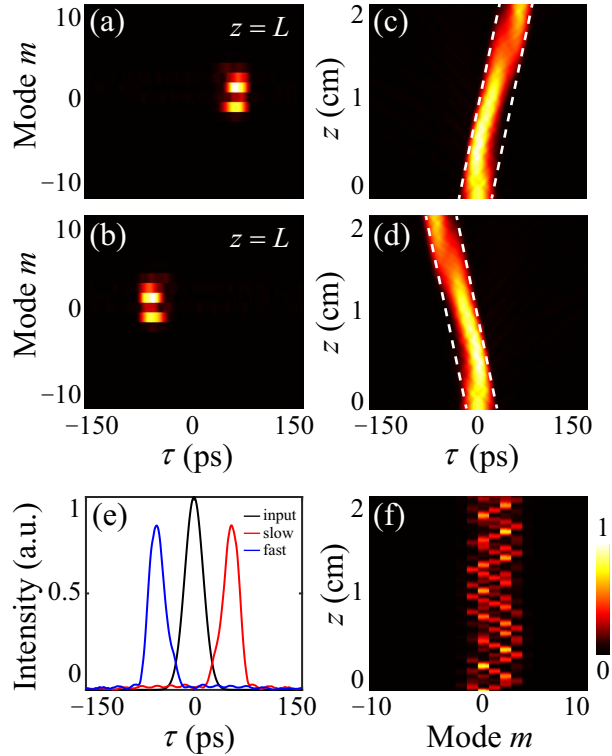}
\caption{\label{fig3}   (a)-(b) Intensity distribution $|a_m(z,\tau)|^2$ at  $z=L$ in the synthetic time-frequency space.   (c)-(d)  Evolution of the pulse $I(z,\tau)$. Effective gauge potential wells used are labeled by dashed lines. (e) The intensity profile of output pulse $I(z=L,\tau)$ in (a)(red line) and (b)(blue line), compared with input pulse $I(z=0,\tau)$ (black line). (f) Evolution of intensities for each mode $I_m(z)$  with effective well in (c). }
\end{figure}

We further shift the  center of the effective gauge potential well  lineally dependent on $z$  with $\tau_\texttt{c}(z)=\eta_1z$, where $\eta_1$ is the shift parameter. This choice makes the modulation phase taking the form of  $\phi(z,\tau)=\pi$ for $|\tau-\eta_1z|\leq 30$ ps and zero for other $\tau$. It provides a way to manipulate the group velocity of pulse controlled by  $\eta_1$. In simulations, we use $\Omega=2\pi\cdot29.4$ GHz, $\Delta n=10^{-3}$, and $k_2=6\times10^{-20}$ m$^{-1}$s$^2$, which give $g=2\times10^3$ m$^{-1}$, and $c_2=1023$ m$^{-1}$.  With a positive  $\eta_1=22$ ps/cm, we find slow  light with group velocity delay of 57 ps and unchanged pulse width of 30 ps as shown in  Figs.~\ref{fig3}(a) and \ref{fig3}(c).  On the other hand, one can see the generation  of fast light in Figs.~\ref{fig3}(b) and \ref{fig3}(d) with negative $\eta_1=-22$ ps/cm. Here large dispersion and strong modulation are chosen to efficiently manipulate the group velocity of the pulse. Figure~\ref{fig3}(e) exhibits the output pulse profiles of slow and fast light at $z=L$, where the peak intensity decreases due to the dispersion loss. The corresponding nonlinear  oscillation for the slow light is plotted in Fig.~\ref{fig3}(f), while the nonlinear  oscillation for the fast light is similar. At $z=L$, the output pulse has two major frequency components at $0^{\texttt{th}}$ and $2^{\texttt{nd}}$  modes [see Figs.~\ref{fig3}(a)-(b)]. The single frequency conversion  can be established by choosing a different propagation length. For example, one sees that oscillations shifts to a single mode at $m=2$ at $z=1.63$ cm. Further larger group velocity manipulation can be obtained by using larger $|\eta_1|$ and GVD at the cost of the intensity loss.

\begin{figure}[htb]
\center
\includegraphics[width=8.0 cm]{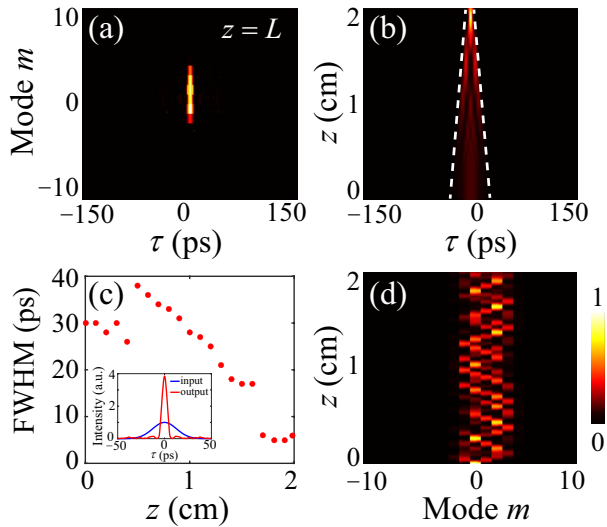}
\caption{\label{fig4}  (a) Intensity distribution $|a_m(z,\tau)|^2$ at  $z=L$ in the synthetic time-frequency space.   (b)  Evolution of the pulse $I(z,\tau)$, where the effective gauge potential wells used are labeled by dashed lines.  (c) Evolution  of the pulse width $\Delta\tau(z)$ calculated from (b). Inserted: the intensity profile of output pulse $I(z=L,\tau)$ in (a)(red line)  compared with input pulse $I(z=0,\tau)$ (blue line).  (d) Evolution of intensities for each mode $I_m(z)$.}
\end{figure}

The width of effective gauge potential well can also be used to control the pulse width. We  consider a varying well along $z$ with $\Delta\tau_{\phi}(z)=30-\eta_2z$ and  $\tau_\texttt{c}(z)=0$,  where $\eta_2$ is the width-varying parameter. Thus  the configuration of modulation phase becomes $\phi(z,\tau)=\pi$ for $|\tau|\leq 30-\eta_2z$ ps and zero for other $\tau$.
We take parameters $\eta_2=12.5$ ps/cm,  $k_2=6\times10^{-21}$ m$^{-1}$s$^2$, $\Delta n=10^{-3}$ and  $\Omega=2\pi\cdot88.2$ GHz, which lead to $g=2\times 10^3$ m$^{-1}$ and $c_2=920$ m$^{-1}$ in the simulation and show results in Fig.~\ref{fig4}. Figure~\ref{fig4}(a) show the pulse profile in the synthetic time-frequency space at $z=L$. Although the pulse converts to multiple modes, the temporal width of the pulse is largely compressed.  As shown in Fig.~\ref{fig4}(b), the narrowing of the effective gauge potential well forces pulse compression gradually while it propagates along the $z$ axis. Figure~\ref{fig4}(c) plots the pulse width versus $z$, showing a trend of overall decrease. The output pulse profile has  width of 5 ps, with the enhanced peak intensity.
Simultaneously, the frequency conversion follows nonlinear  oscillation as shown in Fig.~\ref{fig4}(d). The output with multiple frequency components can be tuned by changing the length of the waveguide. For example, at $z=1.64$ cm, the field exhibits one major frequency component at the $2^{\texttt{nd}}$ mode. Moreover, one  can control the distribution of frequency components in the output pulse by using an input pulse with  a Gaussian distribution of multiple modes. Unidirectional and bidirectional frequency transports together with temporal pulse manipulations can be achieved  \cite{supplementary}. As the last note, the pulse compression can not lead to an output field with an infinitely small temporal width (broad frequency bandwidth). The condition $\Omega>\Delta\omega(z)$ at any $z$ shall be satisfied.

We propose a modulated  waveguide system with parameters based on  lithium-niobate waveguide system. The numbers used in simulations require  waveguide with large dispersion and fast electro-optic modulation, which are experimentally achievable in the second-order nonlinear waveguide with the state-of-art technology. Dispersion near $10^{-22}$ m$^{-1}$s$^2$  has been reported  \cite{8721132}, which can be further enlarged by  structure engineering or operating at higher dispersion wavelength. $\Delta n=10^{-3}$ corresponds to $7$ V/$\mu$m voltage amplitude of an applied external electric field.  If larger modulation is needed,  high  voltage up to  65 V/$\mu$m has been demonstrated \cite{luennemann2003electrooptic}. Shorter pulse manipulation is possible with smaller dispersion but larger modulation frequency, such as 100 GHz, which is commercially available \cite{wang2018integrated}.   Moreover,  recent advances of integrated waveguide platform brings opportunity to achieve synthetic time-frequency dimensions in integrated photonics,  on which  modulators with  frequency $\sim$200 GHz have been demonstrated \cite{liu2020thin}. The  construction of a synthetic time-frequency space can be further extended beyond  by adding other degrees of freedom such as orbital angular momentum \cite{luo2015quantum} or pseudospin \cite{dutt2020a}.  Besides, our analysis shall be valid for other systems if one scales parameters consistently. Our model shows promise for studying pulse propagating in dispersive resonators with synthetic dimensions \cite{shan2020one}, and in third-order dispersive waveguide or microresonator based systems \cite{Hsieh:06,Mia:19,xue2015mode,yang2016broadband}.

In summary, we propose a synthetic (2+1) dimensions for manipulating pulse propagation in a dispersive  waveguide under dynamic modulations. With the effective gauge potential well for photons and  nonlinear oscillation in the synthetic space, multiple pulse propagation behaviors including confined pulse propagation, fast/slow light, and pulse compression have been shown. Our work  provides an alternative platform for actively manipulating single pulse in different ways, which is highly re-configurable, and hence shows promising potentials for pulse engineering  in integrated photonics and optical communications.

\begin{acknowledgments} The research is supported by National Natural Science Foundation of China (12104297, 11974245 and 12122407), National Key R$\&$D Program of China (2017YFA0303701), Shanghai Municipal Science and Technology Major Project (2019SHZDZX01), Natural Science Foundation of Shanghai (19ZR1475700),  and China Postdoctoral Science Foundation (2020M671090). L. Y. acknowledges support from the Program for Professor of Special Appointment (Eastern Scholar) at Shanghai Institutions of Higher Learning.  X. C. also acknowledges the support from Shandong Quancheng Scholarship (00242019024).
\end{acknowledgments}

\bibliography{apssamp}

\end{document}